%% file: main.tex
\documentclass[sigconf, nonacm]{acmart}

\usepackage{multirow}
\usepackage{pifont}
\usepackage{makecell}
\usepackage{xcolor}
\usepackage{enumitem}

\definecolor{darkgreen}{rgb}{0.0, 0.5, 0.0}
\newcommand{\cmark}{\textcolor{darkgreen}{\ding{52}}} 
\newcommand{\xmark}{\textcolor{red}{\ding{56}}} 

\newcommand{\system}{{\textbf{\textsf{\small\textcolor{blue}{ECLAIR}}}}}

\newcommand{\fixed}[1]{\textcolor{black}{#1}}

\begin{document}
\title{Automating the Enterprise with Foundation Models}

\author{Michael Wornow*}
\affiliation{
  \institution{Stanford University}
}
\email{mwornow@stanford.edu}

\author{Avanika Narayan*}
\affiliation{
  \institution{Stanford University}
}
\email{avanikan@stanford.edu}

\author{Krista Opsahl-Ong}
\affiliation{
  \institution{Stanford University}
}
\email{kristaoo@stanford.edu}

\author{Quinn McIntyre}
\affiliation{
  \institution{Stanford University}
}
\email{qam@stanford.edu}

\author{Nigam H. Shah}
\affiliation{
  \institution{Stanford University}
}
\email{nigam@stanford.edu}

\author{Christopher Ré}
\affiliation{
  \institution{Stanford University}
}
\email{chrismre@stanford.edu}

\begin{abstract}
Automating enterprise workflows could unlock \$4 trillion/year in productivity gains. Despite being of interest to the data management community for decades, the ultimate vision of end-to-end workflow automation has remained elusive. Current solutions rely on process mining and robotic process automation (RPA), in which a bot is hard-coded to follow a set of predefined rules for completing a workflow. Through case studies of a hospital and large B2B enterprise, we find that the adoption of RPA has been inhibited by high set-up costs (12-18 months), unreliable execution (60\% initial accuracy), and burdensome maintenance (requiring multiple FTEs). Multimodal foundation models (FMs) such as GPT-4 offer a promising new approach for end-to-end workflow automation given their generalized reasoning and planning abilities. To study these capabilities we propose \system, a system to automate enterprise workflows with minimal human supervision. We conduct initial experiments showing that multimodal FMs can address the limitations of traditional RPA with (1) near-human-level understanding of workflows (93\% accuracy on a workflow understanding task) and (2) instant set-up with minimal technical barrier (based solely on a natural language description of a workflow, \system{} achieves end-to-end completion rates of 40\%). We identify human-AI collaboration, validation, and self-improvement as open challenges, and suggest ways they can be solved with data management techniques. Code is available at: \href{https://github.com/HazyResearch/eclair-agents}{https://github.com/HazyResearch/eclair-agents}

\end{abstract}

\maketitle

\renewcommand{\thefootnote}{\fnsymbol{footnote}} 
\setcounter{footnote}{1} 
\footnotetext{Denotes equal contribution}

\input{1_Intro}
\input{2_Background}
\input{3_Case_Studies}
\input{4_System_Design}
\input{5_Future_Work}

\begin{acks}
We thank Dan Fu, Ben Spector, Vishnu Sarrukai, Jon Saad-Falcon, Simran Arora, Laurel Orr, and Silas Alberti for providing helpful feedback on this manuscript. We thank Ishan Khare, Krrish Chawla, and Miguel Hernandez for their assistance in collecting the dataset of workflow demonstrations. We gratefully acknowledge the support of NIH under No. U54EB020405 (Mobilize), NSF under Nos. CCF2247015 (Hardware-Aware), CCF1763315 (Beyond Sparsity), CCF1563078 (Volume to Velocity), and 1937301 (RTML); US DEVCOM ARL under Nos. W911NF-23-2-0184 (Long-context) and W911NF-21-2-0251 (Interactive Human-AI Teaming); ONR under Nos. N000142312633 (Deep Signal Processing), N000141712266 (Unifying Weak Supervision), N000142012480 (Non-Euclidean Geometry), and N000142012275 (NEPTUNE); Stanford HAI under No. 247183; NXP, Xilinx, LETI-CEA, Intel, IBM, Microsoft, NEC, Toshiba, TSMC, ARM, Hitachi, BASF, Accenture, Ericsson, Qualcomm, Analog Devices, Google Cloud, Salesforce, Total, the HAI-GCP Cloud Credits for Research program,  the Stanford Data Science Initiative (SDSI), and members of the Stanford DAWN project: Facebook, Google, and VMWare. The U.S. Government is authorized to reproduce and distribute reprints for Governmental purposes notwithstanding any copyright notation thereon. Any opinions, findings, and conclusions or recommendations expressed in this material are those of the authors and do not necessarily reflect the views, policies, or endorsements, either expressed or implied, of NIH, ONR, or the U.S. Government. MW is supported by a Stanford HAI Graduate Fellowship and Stanford Healthcare. AN is supported by the Knight-Hennessy Fellowship and the NSF fellowship.
\end{acks}


\bibliographystyle{ACM-Reference-Format}
\bibliography{sample}

\end{document}

%% file: 1_Intro.tex
\section{Introduction}

\begin{figure*}[t]
    \centering
    \includegraphics[scale=0.24]{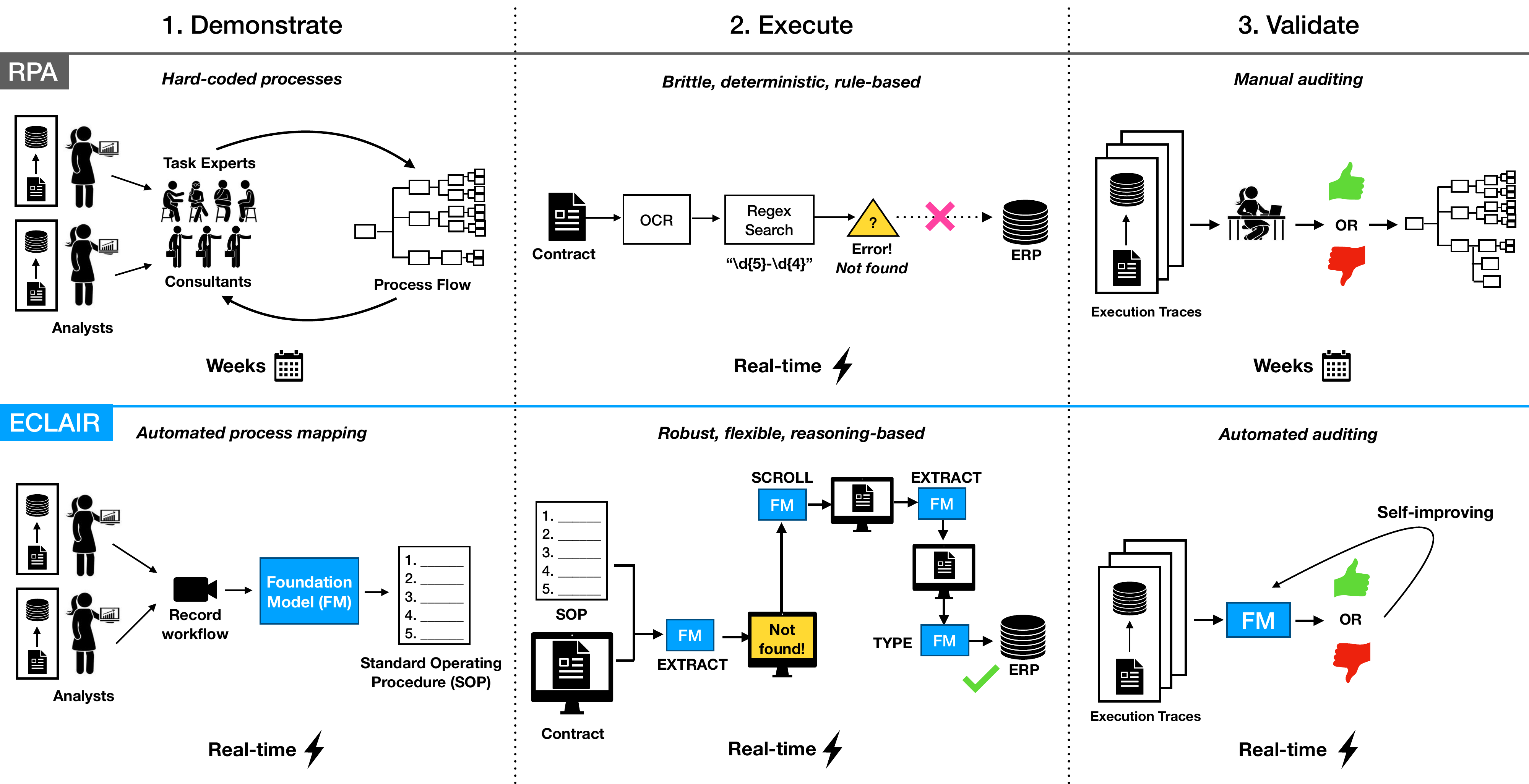}
    \caption{Differences between \system{} and traditional RPA. \fixed{\system{} uses FMs to learn expertise via video demonstrations (left), navigate GUIs given written documentation (center), and audit completed workflows (right).}}
    \label{fig:banner_figure_latest}
\end{figure*}

Digital workflows are the core of our modern economy, with 92\% of jobs now requiring digital skills \cite{bergson2023closing}. Many workflows can and should be automated --- across industries, workers average 3 hrs/day doing repetitive digital workflows tangential to their core jobs \cite{Automation_Anywhere_2020}, a phenomenon referred to as "death by 1,000 clicks" \cite{schulte2019death}. 

For example, consider a large B2B enterprise with tens of thousands of customers (full case study in Section \ref{section:fpa_case_study}). A key accounting workflow is ingesting customer contracts into a centralized database. After receiving a contract via email, an analyst must manually extract relevant data and enter it into an enterprise resource planning system (e.g. NetSuite, SAP, etc.) before they can conduct downstream analyses. Doing this at enterprise scale requires hundreds of individuals, each potentially taking 40 minutes per contract.

As illustrated above, most enterprise workflows involve data integration, ingestion, and transformation, and the market for process management is projected to reach \$65 billion by 2032 \cite{fahland2024well}. \textbf{Decades of data management work has gone into understanding and managing these workflows \cite{casati2000process, sayal2002business, zeng2001agflow, hull2013data, georgakopoulos1995overview, jennings1998adept}, yet the ultimate vision of end-to-end automation --- from understanding to execution to monitoring --- has remained elusive.}

The current best-in-class solution is Robotic Process Automation (RPA), in which workflows identified via process mining get manually encoded into a fixed set of rules for a program to follow \cite{leno2021robotic, moreira2023process, muthusamy2023towards}. Narrowly scoped RPA deployments can have ROIs of 30-200\% and 2x the speed of workflows \cite{lhuer2016next, da2023intelligent}. However, more widespread adoption of RPA has been limited by three key failure modes \cite{sarilo2021slow} which surfaced in interviews with technology leaders at a hospital and B2B enterprise (full case studies in Section 3):

\vspace{-0.5mm}
\begin{itemize}
    \item \textbf{High set-up costs}: The desired workflow must be \textbf{\textit{demonstrated}} to the RPA bot. This is costly, as it requires a trained specialist to map workflows, write automation scripts, and integrate with IT infrastructure \cite{moreira2023process, perdana2023prototyping, hull2013data}. A leading RPA vendor estimates that 3-6 months of experience is needed to become proficient in RPA \cite{UIPath}. In our B2B case study, it took 12 months to go from project kickoff to deployment.
    \item \textbf{Brittle execution}: The RPA bot must \textbf{\textit{execute}} the workflow. Since RPA relies on hard-coded rules, bots cannot adapt to slight variations in input (e.g., a button changing location on a screen, or a form field being renamed) \cite{ye2023proagent, wewerka2020robotic, chakraborti2020robotic}. This leads to a ``death by a thousand cuts'' as the space of possibilities is essentially unbounded. In our B2B case study, the RPA bot was initially only 60\% accurate and took 6 months of improvement to reach 95\%.
    \item \textbf{Burdensome maintenance}: RPA deployments often require human oversight to \textbf{\textit{validate}} outputs and fix edge cases. In our B2B case study, the bot required continual monitoring by \fixed{2 full-time equivalents (FTEs)}.
\end{itemize}

The common cause behind these shortcomings is the difficulty of encoding "tacit" (i.e. difficult to define) human workflow expertise into a rule-based system like RPA \cite{muthusamy2023towards, autor2014polanyi, brynjolfsson2023generative}. Automating most workflows requires planning a sequence of actions and executing them on a graphical user interface (GUI). This requires (a) visual understanding, e.g. identifying a button on a screen; (b) real-time decision making, e.g. knowing to scroll to locate a form field that has shifted location; and (c) common sense to error correct, e.g. hitting escape when an irrelevant pop-up appears.

Multimodal foundation models (FMs) such as GPT-4 \cite{gpt4} have demonstrated visual understanding \cite{fuyu8b, cogvlm, zhang2023visionlanguage} and generalized reasoning abilities \cite{yao2022react, wei2022chain, saycan, voyager} for automating simple digital workflows \cite{seeact, webagent, appagent, mmnavigator_gpt4v_in_wonderland, zhang2024ufo, wu2024copilot, cogagent}. This offers the possibility of sidestepping the failure modes of traditional RPA, just as deep learning eclipsed rule-based approaches over the past decade in machine learning. We thus ask the question: \textbf{Can multimodal foundation models automate enterprise workflows?}


We take a first natural step in studying the opportunities and challenges of applying multimodal FMs across all three stages of traditional RPA by proposing \system{} -- “\textcolor{blue}{\textbf{E}}nterprise s\textcolor{blue}{\textbf{C}}a\textcolor{blue}{\textbf{L}}e \textcolor{blue}{\textbf{AI}} for wo\textcolor{blue}{\textbf{R}}kflows”. As shown in \fixed{Figure \ref{fig:banner_figure_latest}}, our system is defined as follows:

\vspace{-1mm}

\begin{enumerate}
    \item \textbf{Demonstrate:} \system{} uses multimodal FMs to learn from human workflow expertise by watching video demonstrations and reading written documentation. This \textit{lowers set-up costs} and \textit{technical barriers to entry}. Initial experiments show that \system{} can identify every step of a workflow based on screenshots from a demonstration with 93\% accuracy.
    \item \textbf{Execute:} \system{} observes the state of the GUI and plans actions by leveraging the reasoning and visual understanding abilities of FMs \cite{yao2022react, fuyu8b, gpt4}. Based solely on written documentation of a workflow, \system{} improves end-to-end completion rates over an existing GPT-4 baseline from 0\% to 40\% on a sample of 30 web navigation tasks \cite{webarena}. However, this is still far from the accuracy needed for enterprise settings, and we identify opportunities to close this gap.
    \item \textbf{Validate:} \system{} utilizes FMs to self-monitor and error correct. This \textit{reduces the need for human oversight}. When classifying whether a workflow was successfully completed, \system{} achieves a precision of 90\% and recall of 84\%.
\end{enumerate}

Our initial evaluations also identify several patterned failure modes for future research. In \textbf{Execute}, \system{} has difficulty decomposing higher-level steps into discrete actions (e.g. breaking \textit{"Search XXX"} into the sequence of \textit{"click", "type XXX", "press enter"}) and grounding actions to specific GUI elements (e.g. differentiating two buttons with the same label). In \textbf{Validate}, \system{}'s lack of heuristics for navigating GUIs make step-level validation challenging (e.g. checking that a text field is first focused before typing).

While there is still progress to make, we are excited by the potential of \system{} to automate entirely new categories of workflows that require real-time decision-making, interaction with GUIs, and "tacit" domain knowledge \cite{autor2014polanyi}, \fixed{as outlined in Figure \ref{fig:figure_2}}. McKinsey estimates this could double the amount of knowledge work that can be automated \cite{chui2023economic}. 

The rest of the paper is structured as follows. In \textbf{Section \ref{section:background}}, we discuss related work on process mining, RPA, and applying FMs to workflow automation. In \textbf{Section \ref{section:case_studies}}, we provide case studies of a hospital and large B2B enterprise which highlight the limitations of RPA. In \textbf{Section \ref{section:system}}, we outline how \system{} can address these shortcomings. We conclude in \textbf{Section \ref{section:future_work}} with a discussion of \fixed{future work and opportunities for the data management community}. 

Our contributions are: (1) two case studies highlighting the limitations of process mining / RPA, (2) a framework, \system{}, for achieving end-to-end enterprise workflow automation with multimodal FMs, (3) evaluations of \system{} on 30 workflows involving enterprise web applications, and (4) proposals for applying data management techniques to workflow automation.

%% file: 2_Background.tex
\section{Background}
\label{section:background}

We survey related work on RPA, process mining, data management tools for workflow automation, and foundation models (FMs), then detail the problem setting that \system{} seeks to solve.

\begin{figure*}
    \centering
    \includegraphics[scale=0.33]{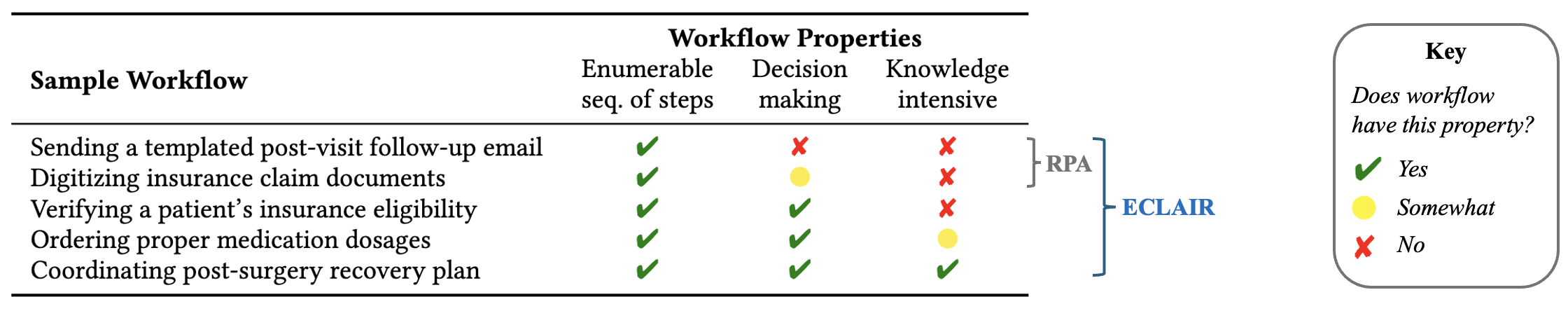}
    \caption{\system{} can automate entirely new categories of workflows, \fixed{such as those that contain hard-to-describe steps, require complex decision making, or are knowledge intensive. Listed examples are real-world hospital workflows (see Section 3.1). }}
    \label{fig:figure_2}
\end{figure*}


\subsection{Related Work}

Significant effort has gone into developing tools for understanding and automating workflows.

 \textbf{\textit{Foundation Models (FMs)}} are deep learning models trained on large datasets which can be adapted to a broad range of downstream tasks \cite{bommasani2021opportunities}. They have demonstrated robust world knowledge \cite{safavi2021relational, jiang2020can}, reasoning \cite{yao2022react, wei2022chain}, and planning abilities \cite{saycan, shen2023hugginggpt, di2023towards, voyager}, and have achieved state-of-the-art results on data processing tasks such as integration and cleaning ~\citep{narayan2022can,kayali2023chorus}.

\textbf{\textit{Process Mining}} is the identification and improvement of workflows based on observational data \cite{van2014process, reinkemeyer2020process, augusto2018automated}. Recent works applied FMs to process mining tasks such as Petri net generation, workflow understanding, and process improvement \cite{fahland2024well, rizk2023case, dumas2023ai, vidgof2023large, berti2023leveraging, grohs2023large, muthusamy2023towards}, but were limited to small case studies and unimodal models.

\textbf{\textit{Robotic Process Automation (RPA)}} is the leading approach for automating enterprise workflows \cite{dahabiyeh2023challenges, ivanvcic2019robotic}. In RPA, a human manually defines a set of rules that a bot then follows to accomplish a specific workflow \cite{ivanvcic2019robotic, chakraborti2020robotic, wewerka2020robotic}. These fixed rulesets make RPA brittle (e.g. failing if a form changes the ordering of its fields) and difficult to maintain \cite{fernandez2021challenges, moreira2023process}. FMs offer a compelling alternative due to their robust reasoning capabilities \cite{yao2022react, saycan} and ability to navigate GUIs \cite{cogagent}. Initial FM-based approaches \cite{mind2web, webagent, bolaa} utilized Large Language Models (LLMs) to act on websites. Since LLMs can only understand text, these works relied on scraping a webpage's HTML as input to the model. This prevented their application to native desktop and virtualized software. Multimodal FMs address this limitation by attaching a vision model to the base LLM \cite{zhang2023visionlanguage}, which enables them to directly reason over screenshots of a GUI \cite{ccnet, webgum, pix2act, cogagent, fuyu8b}. Multimodal FMs have already shown promise in navigating websites  \cite{seeact, appagent, webvoyager}, mobile apps \cite{mmnavigator_gpt4v_in_wonderland}, and desktop applications \cite{wu2024copilot, zhang2024ufo}. We aim to design a system that helps bridge the gap between these proof-of-concepts and enterprise-level solutions.

\textbf{\textit{Data Management for Workflow Automation}} has been studied for nearly two decades, with works ranging from business process management~\cite{zeng2001agflow, sayal2002business, georgakopoulos1995overview, casati2000process, rahman2015worker} to workflow automation and understanding~\cite{hull2013data, jennings1998adept}. All of this work, however, pre-dates multimodal FMs. As a result, the challenges around which these systems were designed differ substantially from modern systems. Our work aims to build on these prior efforts from the data management community by developing a system that integrates FMs.

\vspace{-2mm}

\subsection{Problem Formulation}

We aim to achieve end-to-end automation of enterprise workflows at minimal cost. We have a workflow $w \in \mathcal{W}$ which consists of a sequence of alternating states $s \in \mathcal{S}$ and actions $a \in \mathcal{A}$, such that $w = (s, a, s', a'', ...)$. Each workflow $w$ is done by a set of workers during the course of business operations. These workers follow a standard operating procedure ("SOP"), a form of written documentation which outlines all of the steps and actions of the workflow. Once the workflows are executed, an auditing process validates (either manually or programmatically) whether each workflow $w$ was completed successfully. The goal is to automate workflow $w$ by learning from human demonstrations and written documentation (e.g. SOPs). We aim to directly operate on the GUI, as many workflows do not have APIs or must be executed in native desktop/virtualized applications that can only be observed visually \cite{cogagent}. This necessitates a vision-based approach \cite{seeact, assouel2023unsolved, zhang2024ufo}.

%% file: 3_Case_Studies.tex
\section{Case Studies}
\label{section:case_studies}

We conducted interviews with business leaders across a number of industries who led RPA projects. We select two organizations --- a hospital and large B2B enterprise --- to serve as case studies. Both took over a year to deploy RPA pilot projects, and both declined to expand RPA to additional workflows due to its high set-up costs (\$100k's and months of development), unreliable execution (accuracies started around 60\% and peaked at 95\%), and maintenance requirements. We detail these challenges below.

\vspace{-2mm}
\subsection{Hospital Revenue Cycle Management (RCM)}
\label{section:rcm_case_study}

Given the complexity of healthcare reimbursement, most hospitals have a dedicated Revenue Cycle Management (RCM) department to ensure that timely payment is collected for services delivered \cite{kilanko2023leveraging}. Example RCM workflows include verifying a patient's insurance eligibility, obtaining prior authorization, and processing claims \cite{kilanko2023leveraging}. It is estimated that RCM processes cost roughly 15\% of every dollar of revenue gained \cite{bayley2006hospital}. Additionally, 90\% of health systems face RCM staffing shortages \cite{rcmtrendsr1}. Despite increased interest in automating RCM workflows, most remain highly manual: roughly 94\% of claims submissions and 76\% of eligibility verifications involve manual labor \cite{holloway2018revenue}. In conversations with hospital IT leaders who considered automating certain RCM workflows, but only managed a limited deployment, the following weaknesses of RPA surfaced:

\textbf{\textit{(1) High Set-Up Costs.}} Integration with a hospital's IT infrastructure is costly, as is retraining staff  \cite{kilanko2023leveraging}. Hospital leadership estimated that it took about 18 months and \$10k's to develop and deploy their RPA bot. Significant back-and-forth with the vendor occurred as each workflow had to be manually mapped and coded into a set of well-defined, "always true" actions.

\textbf{\textit{(2) Brittle Execution.}} It was estimated that $\sim$0.2 FTEs worth of effort was saved with the RPA bot, as it could only handle two narrowly-scoped workflows involving one payer and one department. Quarterly updates to the hospital's electronic health record and constant changes to payers' websites would break the bot. Eventually, this required the hospital to develop a custom API that the bot could use to increase its reliability.

\textbf{\textit{(3) Burdensome Maintenance.}} Given staffing constraints, the hospital chose to outsource continued human oversight of the bot to the vendor as a managed service for a fee. Despite this outsourced management, RCM managers still had to manually review outputs to ensure compliance.

\subsection{B2B Enterprise Invoice Processing}
\label{section:fpa_case_study}

Large B2B enterprises must ingest a wide range of complex contracts and process the information into systems of record such as NetSuite or SAP. These workflows can vary widely depending on the specific product being invoiced, the end purchaser, and the time the contract was written \cite{sahu2020invoice}. Given its importance to enterprise financial planning, invoice processing has also attracted attention from RPA vendors. We interviewed the head of revenue operations at a large B2B enterprise which tried to implement RPA for a single invoice processing workflow. After a year of development, 5 FTEs working with the RPA bot were able to successfully accomplish a workflow that previously took 20 FTEs. Despite this apparent success, however, the enterprise chose not to expand RPA to other workflows, as the implementation proved to be too painful:

\textbf{\textit{(1) High Set-Up Costs.}} Going from initial contract with the RPA vendor (\$150k) to production deployment took over 12 months. External consultants (another \$100k) and 3 FTEs were required to integrate the bot with existing IT systems. Major challenges were (a) accurately defining the workflow; and (b) the steep learning curve for programming RPA bots, which required learning a proprietary low-code development toolkit.

\textbf{\textit{(2) Brittle Execution.}} The RPA bot was initially only 60\% accurate post-deployment and took 6 months of iteration to reach a final accuracy of 95\%. Beyond invoice processing, only $\sim$50\% of desired workflows seemed feasible to automate with RPA.

\textbf{\textit{(3) Burdensome Maintenance.}} The cost of a wrongly processed invoice is high (\$10k's), so 2 FTEs were allocated to continuously monitor the bot to debug issues, add new input formats, and inspect outputs. An outside firm also conducted manual batch reviews.

%% file: 4_System_Design.tex
\section{Preliminary Evaluations}
\label{section:system}

In this section, we describe how \system{} can leverage multimodal FMs across all three stages of the traditional RPA pipeline. 

We provide preliminary experiments assessing the feasibility of this approach. For our evaluations, we subsample 30 workflows from the WebArena benchmark \cite{webarena}. WebArena provides a set of interactive websites in which an AI agent must complete complex workflows specified via natural language. Specifically, we choose 30 workflows from the Gitlab and Adobe Magento environments which their GPT-4 baseline model failed to complete \cite{webagent}. We have human annotators record themselves completing each workflow and write a step-by-step guide ("SOP") on the steps they took.

\subsection{Demonstrate}
\label{section:demonstrate}

\system{} aims to learn from passively collected human demonstrations, with no updates to the underlying FM's weights. This limits the cost of labeling data, simplifies deployment, and avoids known biases that arise when humans try to articulate their work processes \cite{li2019interactive}. Our experiments show that GPT-4 can accurately identify the steps of a workflow based on visual observation of a human demonstration, with step-level precision of 0.94 and recall of 0.95.

\subsubsection{Can \system{} determine the steps of a workflow by viewing raw video demonstrations?} This would enable \system{} to substantially improve the effectiveness and scalability of process mining.

\textbf{\textit{Hypothesis:}} A mulitmodal FM can generate accurate SOPs based on screenshots taken at key frames from a video recording.

\textbf{\textit{Set-Up:}} We prompt GPT-4 to generate an SOP for a workflow given a human demonstration. We ablate various ways to provide the demonstration: just the workflow description (WD); the workflow description and screenshots of key frames from a video recording of the demonstration (WD+KF); or the workflow description, key frames, and a textual action log of each click and keystroke (WD+KF+ACT). Using a manually-written SOP as reference, a human annotator calculates GPT-4's precision (\textit{"What percent of steps in the GPT-4 SOP are in the true SOP?"}), recall (\textit{"What percent of steps in the true SOP are in the GPT-4 SOP?"}), and correctness (\textit{"By following the GPT-4 SOP, can I complete the workflow?"}). 

\begin{table}[h]
    \footnotesize
    \caption{(Demonstrate) GPT-4 generation of SOPs. Metrics averaged across all 30 workflows.}
    \begin{tabular}{ccccccc}
        \toprule
        \multirow{2}{*}{\textbf{Method}} & \multicolumn{3}{c}{\textbf{\# of Steps in SOP}} & \multicolumn{3}{c}{\textbf{Accuracy of SOP}}  \\
        & Missing & Incorrect & Total & Precision & Recall & Correctness \\
        \midrule
        WD & 1.57 & 3.58 & 13.67   & 0.75 & 0.81 & 0.60\\
        WD+KF  & 0.67 & 1.05 & 10.17 & 0.89 & 0.92 & 0.90 \\
        WD+KF+ACT  & 0.63 & 0.57 & 9.63 & \textbf{0.94} & \textbf{0.95} & \textbf{0.93}  \\
        \hline
        Ground truth & 0 & 0 & 8.70 & 1 & 1 & 1 \\
        \bottomrule
    \end{tabular}
    \label{tab:demonstrate}
\end{table}

\textbf{\textit{Takeaways:}} As shown in Table \ref{tab:demonstrate}, the SOPs generated by GPT-4 using the WD+KF+ACT strategy are judged as sufficiently "correct" to complete 93\% of workflows. Even providing GPT-4 with screenshots alone (WD+KF) achieves 90\% correctness. However, WD+KF experiences almost twice as many hallucinations (1.05 incorrect steps per SOP versus only 0.57 when the action trace is included). Note that we do not do any workflow-specific prompt engineering or data labeling to achieve these results. \fixed{Additionally, we preprocess our video demonstrations into a sequence of key frames using imperfect heuristics (i.e. alignment with clicks and keystrokes). Future work may benefit from using a model that can directly process video (rather than just images).}

\subsection{Execute}
\label{section:execute}

After defining the workflow, \system{} must execute a sequence of steps to accomplish the workflow. We divide each step into two phases: (1) action suggestion, i.e.  planning what action to take; (2) action grounding, i.e. translating the plan into actual clicks/keystrokes of GUI elements. We find that providing domain knowledge to \system{} via SOPs doubles workflow completion rates, while general-purpose models such as GPT-4 lag smaller models fine-tuned on GUIs for action grounding.

\vspace{-1mm}

\subsubsection{Can \system{} accurately suggest the next action to take in a workflow?} We investigate whether multimodal FMs can predict the next action in a workflow based on the current state of the GUI and action history. We evaluate if providing the FM with an SOP for the workflow increases overall workflow completion rates.

\textbf{\textit{Hypothesis:}} Providing high-level natural language guidance to the model via an SOP improves workflow completion rates. 

\textbf{\textit{Set-Up:}} At each step of the workflow, the model takes as input the ground truth history of actions, full SOP, and current GUI, and is expected to generate the next action to take. We measure the accuracy of the model's suggested next action using a human annotator to evaluate whether it is semantically equivalent to the corresponding "ground truth" action for the workflow.

\vspace{-1mm}

\begin{table}[h]
  \footnotesize
  \caption{(Execute) GPT-4 average accuracy on next action suggestion with and without SOP guidance.}
  \begin{tabular}{ccccc}
    \toprule
    SOP & \makecell{Next Action\\Suggestion Acc.}  & \makecell{Overall Workflow\\Completion Acc.} \\
    \midrule
    \xmark &  0.83 & 0.17 \\
    \cmark & \textbf{0.92} & \textbf{0.40}  \\
    \bottomrule
  \end{tabular}
  \label{tab:execute_sop}
\end{table}\vspace{-0.2cm}

\textbf{\textit{Takeaways:}} Our results in Table \ref{tab:execute_sop} demonstrate that SOPs improve overall workflow completion rates by up to 23 points. While the SOP boosts the accuracy of each step's action suggestion to 0.92, the model struggles to associate these suggested actions with the appropriate GUI elements, thereby deflating its overall completion rate (recall that the model must ground its actions to successfully complete the workflow). For example, when the action is \textit{``Click on the profile button"}, but the HTML element for the icon is identified as "svg" rather than "button", the model fails to select this element.

\subsubsection{Can \system{} accurately "ground" its actions suggestions to GUI elements?} Once \system{} suggests an action to take (i.e. \textit{``Click on the "Submit" button"}), it must then map the action into mouse/keyboard commands which specify the pixel location of the GUI element to interact with \cite{seeact}. 
Multimodal FMs are known to have difficulty with this, so we evaluate several grounding strategies \cite{seeact}.

\textbf{\textit{Hypothesis:}} Providing bounding boxes for each element in a GUI improves a multimodal FM's ability to ground elements.

\textbf{\textit{Set-Up:}} We sample a total of 120 and 302 webpages from the WebUI \cite{webui} and Mind2Web \cite{mind2web} datasets, respectively. We create a natural language description for one element on each page, then prompt a model to generate a bounding box (BB) for that element given a natural language description of the \fixed{BB} and screenshot of the webpage. We measure "accuracy" as the percentage of predicted BBs whose center is within the element's true BB (i.e. \textit{If the model clicked on the center of its prediction, would it successfully hit the target element?}). We evaluate GPT-4 \cite{gpt4} and CogAgent \cite{cogagent} as two state-of-the-art closed/open source multimodal FMs for GUI navigation. While CogAgent directly outputs BBs, GPT-4 does not. Thus, we use "set-of-marks" prompting for GPT-4, in which we overlay a unique numeric label on top of every element in the webpage screenshot provided to GPT-4, and have it output the number of a labeled element \cite{setofmarks}. We generate these labels either directly from the webpage's HTML ("HTML") or from a YOLONAS object detection model ("YOLO") finetuned on 7k WebUI webpages \cite{webui}. The latter simulates the setting where HTML is not available.

\begin{table}[h]
  \caption{(Execute) Accuracy on grounding actions to GUI elements. "S | M | L" is accuracy on small, medium, and large elements.
  \textit{Note: We found that HTML bounding boxes were not accurate in Mind2Web, so it is excluded.}}
  \footnotesize
  \centering
    \begin{tabular}{cccccc}
    \hline
    \multirow{2}{*}{\textbf{Model}} & \multirow{2}{*}{\textbf{\makecell{Bbox\\Source}}} & \multicolumn{2}{c}{\textbf{Mind2Web}} & \multicolumn{2}{c}{\textbf{WebUI}} \\
     & & S | M | L & Overall & S | M | L & Overall \\
    \hline
    GPT-4 & -- & 0.01 | 0.03 | 0.16 & 0.07 & 0.00 | 0.03 | 0.14 & 0.05 \\
    GPT-4 & YOLO & 0.38 | 0.68 | 0.80 & 0.62 & 0.50 | 0.58 | 0.69 & 0.58 \\
    GPT-4 & HTML & -- & -- & 0.56 | 0.58 | 0.67 & 0.60 \\
    CogAgent & -- & \textbf{0.55} | \textbf{0.71} | \textbf{0.87} & \textbf{0.71} & \textbf{0.71} | \textbf{0.64} | \textbf{0.75} & \textbf{0.70} \\
    \hline
    \end{tabular}
    \label{tab:grounding}
\end{table}

\textbf{\textit{Takeaways:}} The 18-billion parameter CogAgent outperforms GPT-4 on action grounding. This suggests that smaller models purpose-built for GUI navigation can be more effective than larger, general purpose multimodal FMs. However, overall accuracy peaks at 70\%, and interacting with smaller-sized elements remains challenging. For GPT-4, using the bounding boxes generated by the YOLONAS model performs similarly to the "ground truth" HTML boxes. This suggests that detecting elements on a GUI with a vision model is not the bottleneck, but rather choosing which of those detected elements is the desired element to interact with.

\subsection{Validate}
\label{section:validate}

There are several levels of validation that \system{} must provide. At the individual step-level, the agent should validate that its action suggestions are (a) feasible to execute and (b) making progress towards accomplishing the workflow. At the workflow-level, the agent should understand (a) whether the workflow was successfully completed and (b) whether the steps taken to achieve the workflow were sensible. We find that GPT-4 struggles with the former lower-level details but performs well at the latter higher-level reasoning.

\subsubsection{Can \system{} self-monitor at the individual step-level?} \fixed{Identifying errors in individual actions would allow ECLAIR to perform error correction in real-time. This would improve the reliability of the Execution stage by enabling the model to avoid undesirable states and backtrack where appropriate.}
\label{section:validate_step_level}

\textbf{\textit{Hypothesis:}} A multimodal FM can detect if an action will succeed or fail based on visual observation of changes in screen state.

\textbf{\textit{Set-Up:}} First, we test the model's ability to detect if an action failed (e.g. typing had no effect because no text field was first focused). We sample $(s, a, s')$ traces from our dataset (positive examples) and generate tuples where $s' = s$ (negatives). We prompt GPT-4 to identify if the action $a$ in $(s, a, s')$ was successfully executed, given screen shots for $s$ and $s'$. We sample three negatives for each positive example. Second, inspired by prior work on data cleaning \cite{benedikt2015querying, baltopoulos2011maintaining}, we create a set of "integrity constraints" defining whether an action is viable at a particular state. For example, an "integrity constraint" for clicking a button is that the button is visible and not disabled. We annotate constraints for all actions in our dataset, then prompt GPT-4 with $(c, s)$ pairs where $c$ is the constraint for the action directly after state $s$ (positive examples) and  $(c, s')$ pairs where $s'$ is a random state occurring before $s$ (negatives).

\begin{table}[h]
  \caption{(Validate) Performance of GPT-4 on self-validation tasks. Metrics averaged across all 30 workflows.}
  \footnotesize
  \begin{tabular}{cccc}
    \toprule
    \textbf{Eval Type} & \textbf{Precision} & \textbf{Recall} & \textbf{F1}\\
    \midrule
    Actuation & 0.95 & 0.85 & 0.90 \\
    Integrity Constraint & 0.67 & 0.36 & 0.47 \\
    Workflow Completion & 0.90 & 0.84 & 0.87 \\
    Workflow Trajectory & 0.88 & 0.83 & 0.85 \\
    \bottomrule
  \end{tabular}
  \label{tab:validate}
\end{table}

\textbf{\textit{Takeaways:}} The results are shown in rows "Actuation" and "Integrity Constraint" in Table~\ref{tab:validate}. GPT-4 has a high precision (0.95) and recall (0.85) when assessing whether an action was successfully executed. However, the low integrity constraint scores indicate that it struggles to identify which actions are viable given the state of the GUI. This could be due to only observing static screenshots, which makes it difficult to discern animations such as a blinking cursor. These results suggest that current multimodal FMs cannot adequately self-monitor for enterprise use cases.

\subsubsection{Can multimodal FMs self-monitor at the overall workflow-level?} Understanding whether a workflow was successfully completed or not can help the agent (a) at runtime to know when to stop execution, and (b) post-deployment for self-auditing.

\textbf{\textit{Hypothesis:}} A multimodal FM can determine if a demonstration correctly achieved a workflow based on visual observation of changes in screen state.

\textbf{\textit{Set-Up:}} First, we test if the model can determine if it successfully completed a workflow. To evaluate this, we sample full traces of $(s^1, a^1,..., a^{n-1}, s^n)$ from our dataset (positive examples) and truncate some by a random number of frames to get $(s^1, a^1,...,a^{k-1}, s^k)$ (negatives) where $k < n$. Given the trace and workflow description, we prompt GPT-4 to provide a binary assessment of whether the workflow was successfully completed. Second, we investigate whether the model understands the proper \textit{trajectory} of actions in a workflow --- i.e. it is not sufficient to merely complete the workflow, but the steps taken to complete it must align with its SOP. We sample full traces $(s^1, a^1,..., a^{n-1}, s^n)$  from our dataset (positives) and either (a) randomly shuffle or (b) randomly delete frames from this trace (negatives). We provide GPT-4 with the SOP for the workflow, the workflow description, and the trace, and have it output a binary assessment of whether the trace exactly followed the SOP. 

\textbf{\textit{Takeaways:}} The F1 scores of 0.87 and 0.85 in rows "Workflow Completion" and "Workflow Trajectory" of Table~\ref{tab:validate} suggest that GPT-4 can self-monitor higher-level properties of a workflow, but still has significant room for improvement.

%% file: 5_Future_Work.tex
\section{Discussion}
\label{section:future_work}

\fixed{To be deployed, \system{} must meet a minimum level of performance. While this is highly workflow-dependent, \system{} must be accurate enough such that the cost of correcting its errors is outweighed by the efficiency gains of using it (which might not require 100\% accuracy). In our Section \ref{section:fpa_case_study} case study, for example, an accuracy of 95\% was sufficient for the RPA bot to be deployed. Defining such success criteria can be done via interviews with stakeholders, financial modeling, and other enterprise planning methods \cite{leno2021robotic} We discuss considerations for deploying \system{} below.}

\textbf{\textit{Error Handling and Monitoring.}} \fixed{We envision a multi-tiered system which combines (1) self-validation, (2) programmatic heuristics, and (3) limited human intervention to provide error correction. (1) Per our Section \ref{section:validate} experiments, a repository of integrity constraints --- which have successfully enhanced the quality of database schemas \cite{benedikt2015querying, baltopoulos2011maintaining} --- could improve the accuracy of FM self-monitoring. Though FMs can exhibit non-deterministic behavior, their reliability can be improved by setting their temperature to 0, repeatedly querying \cite{shinn2023reflexion} and ensembling predictions \cite{li2024more}, or eliciting confidence scores to surface cases where intervention is necessary \cite{tian2023just}. (2) Programmatic heuristics can also be used to detect failures, e.g. deviation from the average time to execute a workflow. (3) Existing methods for auditing human workflows can be re-applied, such as random spot checks or screenshots of confirmation screens. However, we envision such monitoring to be more limited -- namely, to close the knowledge gap between \system{} and a particular domain -- as the outputs from monitoring can be repurposed into a dataset that can be used to improve \system{}~\cite{ouyang2022training}.}

\textbf{\textit{Human-\system{} Collaboration.}} \fixed{While the long-term vision for \system{} is to require minimal human interaction --- i.e. only the Demonstrate phase (Section \ref{section:demonstrate}) needing human involvement --- we acknowledge that human supervision may be necessary for certain workflows. For example, a physician sign-off before prescribing medications or tasks involving user authentication. To accomplish this, the SOP could mark steps where the model transfers control to a human. Alternatively, a whitelist of sensitive actions can be compiled to automatically force transfer of control to a human when triggered, similar to how kernels use interrupts to handle control flow \cite{mejia2018interrupt}. Finally, as mentioned in the prior section, generated human-\system{} execution traces can be used to improve \system{} performance via fine-tuning or few-shot prompting~\cite{ouyang2022training}.}

\textbf{\textit{Self-Improvement.}}
As \system{} repeatedly executes a workflow, it can observe the effects of its actions on the environment. By documenting these observations, \system{} can compile a database of common "skills" that can later be transferred to different workflows \cite{fu2024drive, appagent, wu2024copilot, wang2023jarvis, park2023generative}. Applying principles from self-driving databases, which aim to continuously improve performance by implementing sequences of "actions" (i.e. changes to their configurations) based on utilization patterns \cite{pavlo2021make, pavlo2017self, ma2018query}, could provide a principled approach for such a self-improving workflow automation system.

\textbf{\textit{Multi-Agent Collaboration.}} 
\fixed{Applying multiple agents to the same task can improve accuracy \cite{li2024more, liang2023encouraging}, as seen in recent work on multi-agent software engineering \cite{hong2023metagpt} and chatbot applications\cite{wu2023autogen}. Such an approach could be utilized within the \system{} framework to create specialized agents for distinct subtasks or digital environments. Prior work on collaborative data processing tools offers a reference for how \system{} can be scaled to multi-agent (and multi-human) workflows with shared resources. \cite{bhardwaj2015collaborative, liu2022demonstration}.}

\vspace{-1mm}

\section{Conclusion}

We are excited for the potential of multimodal FMs to reimagine how work gets done. By addressing the three main shortcomings of traditional process mining and RPA (high set-up costs, brittle execution, and burdensome maintenance), the realization of \system{} can help achieve the promise of enterprise workflow automation.